\documentclass[12pt]{article}
\usepackage{amsmath}
\usepackage{latexsym}
\usepackage{amssymb}
\usepackage{a4wide}
\usepackage{bbm}
\usepackage{epsfig}
\newcommand{\I}{\text{i}}
\newcommand{\E}{\text{e}}

\newcommand{\re}[1]{~(\ref{#1})}\begin{document}
\newcommand{\meff}{m_{\text{eff}}}
\newcommand{\Geff}{\Gamma_{\text{eff}}}
\newcommand{\Leff}{{\cal L}_{\text{eff}}}
\newcommand{\Bcr}{B_{\text{cr}}}
\newcommand{\GF}{G_{\text{F}}}
\newcommand{\gA}{g_{\text{A}}}
\unitlength=1mm
\title{\bf From effective actions to actual effects\\ in
  QED\thanks{Presented at QED 2000, the $2^{\text{nd}}$ Workshop on
    ``\textsc{frontier tests of quantum electrodynamics and physics of
      the vacuum}'', at Trieste,
    Italy, Oct. 5-11, 2000}}
\author{Holger Gies\\
  {\small\it Institut f\"ur theoretische Physik, Universit\"at
    T\"ubingen,}\\{\small \it   72076 T\"ubingen, Germany}}
\maketitle
\begin{abstract}
  The construction of low-energy effective actions in QED for several
  types of external conditions is reviewed. Emphasis is put on the
  application of these effective actions to a variety of physical
  effects which represent a manifestation of vacuum
  polarization. Soft-photon interactions with external electromagnetic 
  fields and/or a heat bath are described, pair production at finite
  temperature is discussed, and finally a glance at photon-neutrino
  interactions is provided.
\end{abstract}

\section{Introduction}
The most disturbing ingredient in quantum electrodynamics (QED) is the
mass of the electron. It separates the fields and their ubiquitous
quantum fluctuations into two classes: those with momenta larger than,
and those with momenta smaller than, the electron mass $m$. Naturally,
electrons can only appear in the first class which gives rise to the
distinctiveness of electromagnetic fields with momenta smaller than
$m$. In this talk, we shall turn our attention to these, so-called
``soft'', fields and their dynamics, and how this dynamics is
influenced by the ``hard'' fluctuations of electrons and photons.
Since $m\hat{=} 7.6\times 10^{11}$GHz, the terminus ``soft field''
covers a considerably wide range of electromagnetic fields.

From the full quantum theory of electrons and photons, we can arrive
at an ``effective'' description of the soft electromagnetic fields in
terms of an effective action; the latter is obtained by averaging over
the hard fluctuations and condensing their influence into a number of
interaction couplings between the soft fields. This program can also
be carried out in the presence of various external perturbations, a
so-called modified vacuum, which affect the high-momentum
fluctuations. After the averaging, the external perturbations are also
reflected in the soft-field couplings and thereby modify the dynamics of
the soft fields indirectly.

\begin{figure}
\begin{center}
\begin{picture}(140,40)
\put(14,0){\epsfig{figure=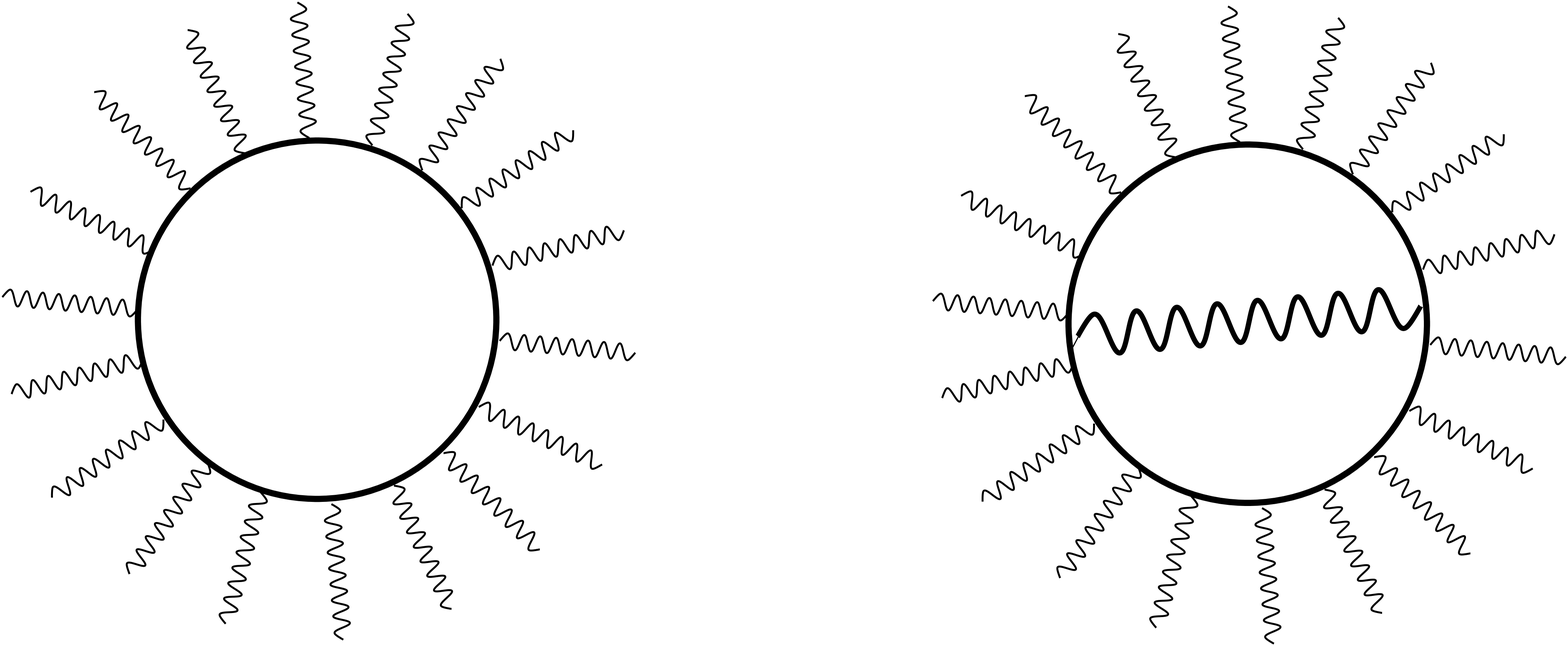,width=12cm}}
\put(0,40){(a)}
\put(70,40){(b)}
\end{picture}
\end{center}

\vspace{-0.8cm}

\caption{(a) Diagrammatic representation of the QED effective action 
   at the one-loop level corresponding to Eq.\re{1}. (b) Two-loop
   level of the effective action as employed in Sect. 2 involving a
   thermalized radiative photon.}
\end{figure}

This talk reviews a personal selection of such effective
actions\footnote{For technical details about the construction of these
  actions, the interested reader is referred to the original
  literature; for a comprehensive collection of large parts of the
  material presented here, see \cite{Dittrich:2000zu}.} adapted to
different physical systems and presents some of their applications to
physical effects.

Let us begin with a brief sketch of the classic example: the
Heisenberg-Euler action \cite{heis}. Here, the physical system is
defined simply by soft electromagnetic fields being placed in the
vacuum. The vacuum, however, is filled with fluctuating electrons and
positrons that couple directly to the external fields. Averaging over
the fermions results in an interacting theory of the soft fields
themselves:
\begin{eqnarray}
\Geff^{\text{HE}}\!\!\!\!&=&\!\!\!\!\int\! d^4x\Biggl\{ -{\cal F}  +
\frac{1}{8\pi^2} \int\limits_0^\infty
  \frac{ds}{s^3} \E^{-\I m^2s} \nonumber\\
&&\!\times\left\{(es)^2|{\cal G}| \cot \bigl[ es(\sqrt{{\cal F}^2
  \!+{\cal G}^2} \!+{\cal F})^{\frac{1}{2}} \bigr] \coth \bigl[
  es(\sqrt{{\cal F}^2\! +{\cal G}^2} \!-{\cal F})^{\frac{1}{2}} \bigr]
  +\frac{2}{3} (es)^2 {\cal F}\! -1 \right\}\!\Biggr\}\nonumber\\
&=&\int d^4x\left( -{\cal F}+\frac{8}{45}
  \frac{\alpha^2}{m^4}\, {\cal F}^2  +\frac{14}{45}
  \frac{\alpha^2}{m^4} \,{\cal G}^2 + {\cal O}(F^6) \right), \label{1}
\end{eqnarray} 
where we introduced the only gauge and Lorentz invariants of the
electromagnetic field
\begin{equation}
{\cal F}:=\frac{1}{4} F_{\mu\nu} F^{\mu\nu} = \frac{1}{2} (\mathbf{B}^2
-\mathbf{E}^2 ), \qquad 
{\cal G}:=\frac{1}{4} F_{\mu\nu} {}^\star F^{\mu\nu} = 
-\mathbf{E\cdot B}. \label{2}
\end{equation}
Equation \re{1} displays the one-loop Heisenberg-Euler action in
proper-time representation as derived by Schwinger
\cite{Schwinger:1951nm} (see Fig.~1(a)); the last line is a weak-field
expansion thereof including the 4-photon-vertex level. The first term
is the classical Maxwell term and the interaction terms $\sim
\frac{\alpha^2}{m^4}$ are given in one-loop approximation. Only
because the electron is so ``heavy'', can the world of human
experience be appropriately approximated by what we call classical
electrodynamics owing to the smallness of the next-to-leading order
terms.

Beyond the classical approximation, a number of effects appear as a
consequence of the self-interactions of the soft fields. In the first
place, we have light-by-light scattering as a direct violation of the
classical superposition principle. Of course, the integrated total
elastic cross section as derived from the last line of Eq.\re{1} is
very small: 
\begin{equation}
\sigma=\frac{973}{10125} \frac{\alpha^2}{\pi} \, r_0^2 
 \left(\! \frac{\omega}{m}\! \right)^6, \label{LG13}
\end{equation}
where $r_0$ is the classical electron radius, and the center-of-mass
frequency of the photon is small by assumption, $\omega\ll m$.

\begin{figure}
\begin{center}
\begin{picture}(140,40)
\put(10,3){\epsfig{figure=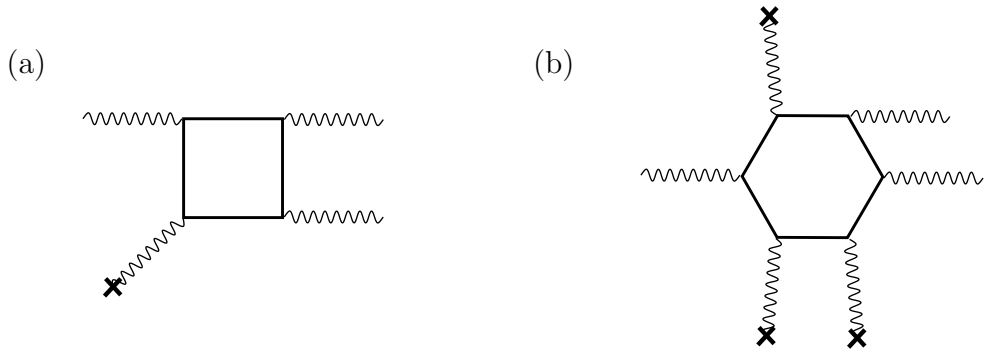,width=12cm}}
\put(0,40){(a)}
\put(70,40){(b)}
\end{picture}
\end{center}

\vspace{-1cm}

\caption{(a) The photon-splitting box graph is suppressed at zero
  temperature, but becomes dominant at finite temperature
  (cf. Eq.\re{PS20}). (b) The hexagon graph dominates zero-temperature
  photon splitting (cf. Eq.\re{4}). Crosses denote couplings to the
  magnetic background field.} 
\end{figure}

As a second example, we may consider two different types of soft
fields, e.g., a propagating soft photon in the background of an
electromagnetic field. As was shown by Adler \cite{Adler:1971wn}, such
a photon can decay into two photons in the presence of a background
magnetic field. This photon splitting can act as a production
mechanism of polarized light near strongly magnetized astrophysical
objects because the splitting of photons with perpendicular
polarization WRT the $\mathbf{B}$ field into parallelly polarized
ones is preferred, ($\bot \to \|_1+\|_2$). The absorption coefficient
of this process as derived from Eq.\re{1} is given by
\begin{equation}
\kappa\simeq 0.116 \,\text{cm}^{-1}
\left(\frac{eB}{m^2}\right)^6 \sin^6\theta_B\left
  ( \frac{\omega}{m}\right)^5\!,\quad B\ll \Bcr:= \frac{m^2}{e},\quad
\theta_B=<\!\!\!)\, (\text{prop.\,\,direct.},\mathbf{B}).
\label{4} 
\end{equation}
One peculiarity of this process is the $B^6$ dependence: it indicates
that three couplings of the background field to the $e^+e^-$ loop are
necessary for the lowest-order process. Together with the one incoming
and the two outgoing photons, this constitutes a hexagon graph; the box
graph with one coupling to the $B$ field is suppressed owing to
properties of the Lorentz algebra of the field strength tensor (see
Fig.~2).

The third example is the seeming paradox that light in an
electromagnetic background no longer propagates ``at the speed of
light''. The vacuum modified by the background field acts like a
medium in classical electrodynamics shifting the phase and group
velocities of propagating soft photons, including effects of
birefringence\footnote{The world is currently awaiting the
  observation of the birefringence phenomenon as a first direct
  verification of the effective nonlinearity of electrodynamics
  \cite{peng98}.} 
and dispersion. In many cases, the polarization-averaged 
deformation of the light cone $k^2$ can be described by 
\cite{Dittrich:1998fy}
\begin{equation}
k^2=Q\, \langle T^{\mu\nu}\rangle\, k_\mu k_\nu, \label{5}
\end{equation}
where $\langle T^{\mu\nu}\rangle$ denotes the expectation value of the
energy-momentum tensor in the modified vacuum, and the proportionality
factor $Q$ depends on the parameters of the effective action
(recently, a formulation of Eq.\re{5} in terms of an ``effective
metric'' has been given \cite{Novello:2000pg}). For the case of a
magnetic background field, the classical result as obtained from the
last line of Eq.\re{1} for the velocity of light is
\cite{Adler:1971wn,bial70,Dittrich:1998gt}:
\begin{equation}
v\simeq 1-\frac{11}{45} \frac{\alpha^2}{m^4}\, B^2\,
\sin^2\theta_B. \label{6}
\end{equation}

As a final introductory example, we would like to mention the
Schwinger mechanism for $e^+e^-$ pair production in electric fields,
representing the instability of the vacuum that is modified by an
electric field. The pair production rate per unit volume and time is
related to the imaginary part of the effective action \re{1}:
\begin{equation}
w=2\,\text{Im}\, \Leff^{\text{HE}} = \frac{(eE)^2}{4\pi^3}
\sum_{n=1}^\infty \frac{\exp \left( -\frac{n\pi m^2}{eE}\right)}{n^2},
\qquad \Geff^{\text{HE}}=\int d^4x \Leff^{\text{HE}}. \label{7}
\end{equation}
It is noteworthy that the Eq.\re{7} is a nonperturbative result, in
the sense that it cannot be obtained from an expansion of the action
in terms of the field strength. The Schwinger mechanism has recently
been verified in the famous SLAC E144 experiment \cite{Burke:1997ew}.
Moreover, it has become a building block of various pair-production
models for heavy-ion collisions.

Generalizations to the Heisenberg-Euler effective action have been
investigated for a variety of further modified vacua including
gravitational backgrounds, finite temperature and/or density,
nontrivial boundary conditions (Casimir vacua), and even couplings to
the QCD vacuum. Exemplarily, we consider the presence of a heat bath
in the following section. In Sect. 3, we go one step further and
enlarge the particle content of our system by taking neutrinos into
account. 

\section{QED effective action at finite temperature}
The generalization of the precedingly presented philosophy to finite
temperature represents a challenge from a purely theoretical as well
as a phenomenological viewpoint\footnote{In the literature, this
  problem has been a touchstone for the various finite-temperature
  formalisms; after decades of discussion, real-time formalism
  \cite{Elmfors:1995fw} and imaginary-time formalism
  \cite{Gies:1999vt} have come to congruent results.}.
Phenomenologically, the presence of a heat bath gives rise to thermal
fluctuations of the electrons, positrons and photons in the vacuum.
Contrary to their quantum analogue, these fluctuations are strictly
on-shell (but can couple to off-shell quantum fluctuations) and create
a (neutral) plasma, containing thermally excited (charged) particles.
The latter can exert an influence on the soft fields under
consideration, which renders their dynamics temperature dependent.

Let us first glance over the theoretical aspects. For the formalism,
the correspondence between quantum field theory and statistical
mechanics for thermal equilibrium is employed, leading to propagators
and wave functions that are periodic in imaginary time with period
$\beta=1/T$.  Furthermore, Lorentz invariance is broken explicitly,
since the rest frame of the heat bath is distinguished; as a new
algebraic element, the 4-velocity vector of the heat bath $u^\mu$
appears ($u^\mu=(1,0,0,0)$ in the heat-bath rest frame) and allows for
the construction of a further invariant beside the temperature itself:
\begin{equation}
{\cal E}:=\bigl(u_\alpha F^{\alpha\mu}\bigr) 
  \bigl(u_\beta F^{\beta}{}_\mu\bigr), \qquad {\cal
  E}={\mathbf{E}}^2\,\,\text{in\,\,the\,\,heat-bath\,\,rest\,\,frame}.
  \label{8}
\end{equation}
The invariants ${\cal E},{\cal F},{\cal G}$ are complete on a
classical level, in the sense that they contain the entire information
about the soft fields. On the quantum level, a new aspect appears:
since the periodicity condition of the wave functions has to be
respected, gauge transformations also have to satisfy this
requirement, leading to a restricted class of gauge transformations
obeying periodicity in imaginary time. This makes room for another
gauge invariant quantity:
\begin{equation}
\bar{A}_u({\mathbf{x}})=\frac{1}{\beta} \int\limits_0^\beta d\tau\,
A_u(x^\mu+\I\tau u^\mu), \qquad A_u:=A^\mu u_\mu, \label{11242}
\end{equation}
where $\mathbf{x}$ denotes the components of $x^\mu$ orthogonal to
$u^\mu$. The physical interpretation of $\bar{A}_u$ is that it has to
be identified with a chemical potential: namely,
$e\bar{A}_u(\mathbf{x})$ denotes the electrostatic energy cost for
placing an electron at position $\mathbf{x}$. The gauge field $A_\mu$
therefore contains considerably more information about the system at
finite than at zero temperature. Moreover, the restricted class of
gauge transformations no longer protects the zero-mass of the photon
as stringently as at zero temperature; it is exactly the $\bar{A}_u$
part of the gauge field that acquires a mass by quantum fluctuations
which has to be identified with the Debye screening mass (see below).

Also, the mechanism by which the quantum invariant $\bar{A}_u$ enters
the effective action is interesting: the periodicity condition for the
quantum fields leads to a compactification of the spacetime manifold
in imaginary-time direction -- the manifold becomes cylindrical,
$\mathbbm{R}^3\times S^1$. The virtual $e^+e^-$ loops therefore fall
into different homotopy classes characterized by their number of
windings around the cylinder. This winding number is measured with the
aid of $\bar{A}_u$, so that the $\bar{A}_u$ dependence of the final
effective action is of topological origin. 

Let us now come to the results for the thermal effective action; at
the one-loop level various exact representations are known for
arbitrary values of temperature and field strength
\cite{Elmfors:1995fw,Gies:1999vt,Elmfors:1998ee}. Here, we confine
ourselves to some limiting cases. Naturally, the electron mass scale
distinguishes between a high-, an intermediate- and a low-temperature
domain, $T\gg m,\, T\sim m,\, T\ll m$. 

As a first limit, we consider $T\gg m$ and vanishing background field
strengths, so that the one-loop thermal contribution $\Geff^{1T}$ only
depends on $\bar{A}(u)$ and can be expanded as $\Geff^{1T}=\int
d^4x\Bigl(-\frac{1}{2} \partial_\mu \bar{A}_u\partial^\mu \bar{A}_u
+\frac{m_{\text{eff}}^2}{2} (\bar{A}_u)^2+ {\cal
  O}(\bar{A}_u^4)\Bigr)$. Here, we can read off the effective mass
that determines the Debye screening effect of electric fields:
\begin{equation}
m_{\text{eff}}^2(T)=\frac{(eT)^2}{3}, \qquad T\gg m,\label{112443}
\end{equation}
which is the well-known result found in the literature employing
different methods. 

Particularly interesting is the limiting case of soft background
fields at high temperature $T\gg m$ (setting $\bar{A}_u=0$ for
simplicity). At one-loop order, the thermally fluctuating electrons
move ultra-relativistically so that the scale of the loop process is
no longer set by the electron mass $m$, but by the temperature $T$.
Therefore, in the limit $T\to \infty$, the loop particles become
infinitely heavy and decouple from the theory. Consequently, vacuum
polarization and thereby any nonlinear interaction of the soft fields
is strongly suppressed. The remaining linear terms are
\begin{equation}
\Geff\equiv \Geff^{\text{HE}}+\Geff^{1T}
= \int d^4x\left( -{\cal F}-\frac{2\alpha}{3\pi} {\cal F} \ln
  \frac{T}{m} + \frac{\alpha}{6\pi} {\cal E} \right),\quad T\gg m. 
\label{9} 
\end{equation}
Besides the classical Maxwell term, we find a finite logarithmic
renormalization of the charge running with temperature; this term
modifies, e.g., the Compton amplitude. The last term $\sim {\cal E}$
is responsible for a decelerated light propagation (see below). 

In the opposite limit of low temperature, the thermal one-loop
contribution to be added to Eq.\re{1} reads in a weak-field
approximation (at zero chemical potential) \cite{Elmfors:1998ee}:
\begin{eqnarray}
\Geff^{1T}\!\!\!&\simeq&\!\!\! \int \!\!d^4x\left\{
\frac{\alpha}{\pi} \left[ \frac{1}{6} \sqrt{\frac{2\pi m}{T}} {\cal E}
  + \frac{2}{3} \sqrt{\frac{2\pi T}{m}} {\cal F} \right] \E^{-m/T}
\right.\nonumber\\
&&\,-\alpha^2 \left[ \frac{1}{36}\frac{m^4}{T^4} \sqrt{ \frac{\pi
      T}{2m}} {\cal E}^2 - \frac{\pi}{90} \frac{m^3}{T^3} (8{\cal
    F}{\cal E} +{\cal G}^2) + \frac{1}{45} \frac{m^2}{T^2} (8{\cal
    F}^2 -8{\cal F}{\cal E} +13{\cal G}^2) \right. \nonumber\\
&&\qquad\qquad\left.\left. +\frac{\pi}{30} \frac{m}{T} (4{\cal
      F}^2+7{\cal G}^2) + \frac{4}{45} (4{\cal F}^2+7{\cal
      G}^2)\right] \frac{\E^{-m/T}}{m^4}\right\}, \qquad T\ll  m. 
\label{1124add1}
\end{eqnarray}
Equation\re{1124add1} can serve as a starting point for similar
explorations of physical effects as discussed above for
$\Geff^{\text{HE}}$ at $T=0$; however, we observe that the thermal
one-loop effective action at low temperature $T\ll m$ is exponentially
suppressed by the electron mass; consequently, any actual effect will
be immeasurably small in this limit. The physical reason for this
suppression is obvious: since the electron spectrum exhibits a mass
gap $m$, a heat bath of temperature $T\ll m$ can hardly excite
electronic states to a significant amount.

This situation changes drastically when we take a look at the next
order in perturbation theory, the two-loop level. Here, an additional
radiative photon has to be considered within the loop. But the photon
is massless, so that higher photon states can be thermally excited
even at small values of temperature. This is the reason why we have to
expect that the thermal two-loop contribution to the effective action
(see Fig.~1(b)) exceeds the thermal one-loop part in the
low-temperature limit. This {\em two-loop dominance} connotes an
inversion of the usual loop hierarchy and signifies that the one-loop
approximation represents an inconsistent truncation of the theory in
this temperature domain; herein, at least one radiative photon has to
be taken into account for correct results.

This situation even intensifies for the imaginary part of the
effective action which is related to pair production in electric
fields\footnote{Here it is tacitly assumed that the system with
  electric fields remains close to thermal equilibrium.}. By
construction, the thermal one-loop contribution is purely real (this
is seen directly in the real-time formalism \cite{Elmfors:1995fw}).
Since the one-loop approximation deals only with thermalized
$e^+$ and $e^-$ which are on-shell by construction, there can be no
thermal contribution to the pair-production rate to this order of
calculation; this is because the $e^+e^-$ pair has to go off-shell in
order to ``tunnel'' through the mass gap $2m$. Again at two loop, the
situation is different, because then, the thermalized on-shell
radiative photon can couple to the fermionic off-shell quantum
fluctuations; in this case, a thermal contribution to pair-production
is no longer prohibited. 

These arguments make clear that an understanding of thermal physics of
soft electromagnetic fields requires a two-loop calculation. The
cumbersome calculation is simplified by the fact that only the
internal radiative photon has to be thermalized because thermal
fermion contributions are suppressed by the electron mass for $T\ll
m$. The thermal two-loop contributions finally read in a weak-field
approximation (at zero chemical potential) \cite{Gies:2000vb}:
\begin{eqnarray}
\Geff^{2T}\!\!&=&\!\!\int\!\! d^4x\left\{\frac{44\alpha^2\pi^2}{2025} 
\frac{T^4}{m^4} ({\cal F}+{\cal E})
-\frac{2^6\cdot37 \alpha^3\pi^3}{3^4\cdot5^2\cdot7} \frac{T^4}{m^4}
\frac{{\cal F}({\cal F}+{\cal E})}{m^4}\right. \nonumber\\
&&\left.+\frac{2^{13}\alpha^3\pi^5 }
{3^6\cdot 5\cdot 7^2}
\frac{T^6}{m^6}\bigl(2{\cal F}^2+6{\cal E}{\cal
  F}+3{\cal E}^2-{\cal G}^2\bigr) \frac{1}{m^4}+{\cal
  O}(F^6,(T/m)^8)\right\}\!,\quad\! T\ll m. \label{2Lft75} 
\end{eqnarray}
We observe a power-law dependence on the temperature starting with
$T^4$; therefore, the two-loop contribution indeed always wins out
over the one-loop term for sufficiently low $T$. This completes our
primary picture of the thermal part of the effective action of QED: at
small values of $T$, the two-loop part governs the dynamics of soft
fields; numerically, we find two-loop dominance for $T/m\lesssim
0.05$. Above, the one-loop part reseizes power over the soft
fields in the intermediate-temperature domain $T\sim m$ and
beyond. Only the thermal imaginary part is completely controlled by
the two-loop action. 

Now let us come to the physics contained in
Eqs.\re{2Lft75},\re{1124add1} and \re{9}, beginning with the subject
of light propagation. The light cone condition for thermal vacua (cf.
Eq.\re{5}) has been derived in \cite{Gies:1999xn}; the final results
read \cite{Barton:1990dq,Gies:1999xn,Gies:2000vb}:
\begin{eqnarray}
v_{2\text{-loop}}&=&1-2\frac{44}{2025} \alpha^2\pi^2
\frac{T^4}{m^4}+{\cal O}(T^8/m^8), \qquad T\ll m, \nonumber\\
v_{1\text{-loop}}&\simeq&1-\frac{\alpha}{6}\, \sqrt{\frac{2}{\pi}}\,
\sqrt{\frac{m}{T}}\, \E^{-\frac{m}{T}}, \qquad T\lesssim m, \label{10}\\
v_{1\text{-loop}}&=&1-\frac{\alpha}{6\pi}+{\cal O}(m^2/T^2), \qquad
T\gtrsim m. \nonumber
\end{eqnarray}
We observe that the phase and group velocities of light decrease for
increasing $T$ until the velocity shift reach a maximum of $\delta
v=\alpha/(6\pi)$ at $T$ a bit above $m$ (see Fig.~3). For
$T/m\gtrsim6$, the considerations become meaningless, because the
thermal plasma develops a plasma frequency corresponding to the Debye
screening mass that acts as a cutoff for the low-frequency modes. For
these values of $T$, ``soft photons'' simply no longer exist.

\begin{figure}
\begin{flushleft}
\begin{picture}(145,56)
\put(4,0){
\epsfig{figure=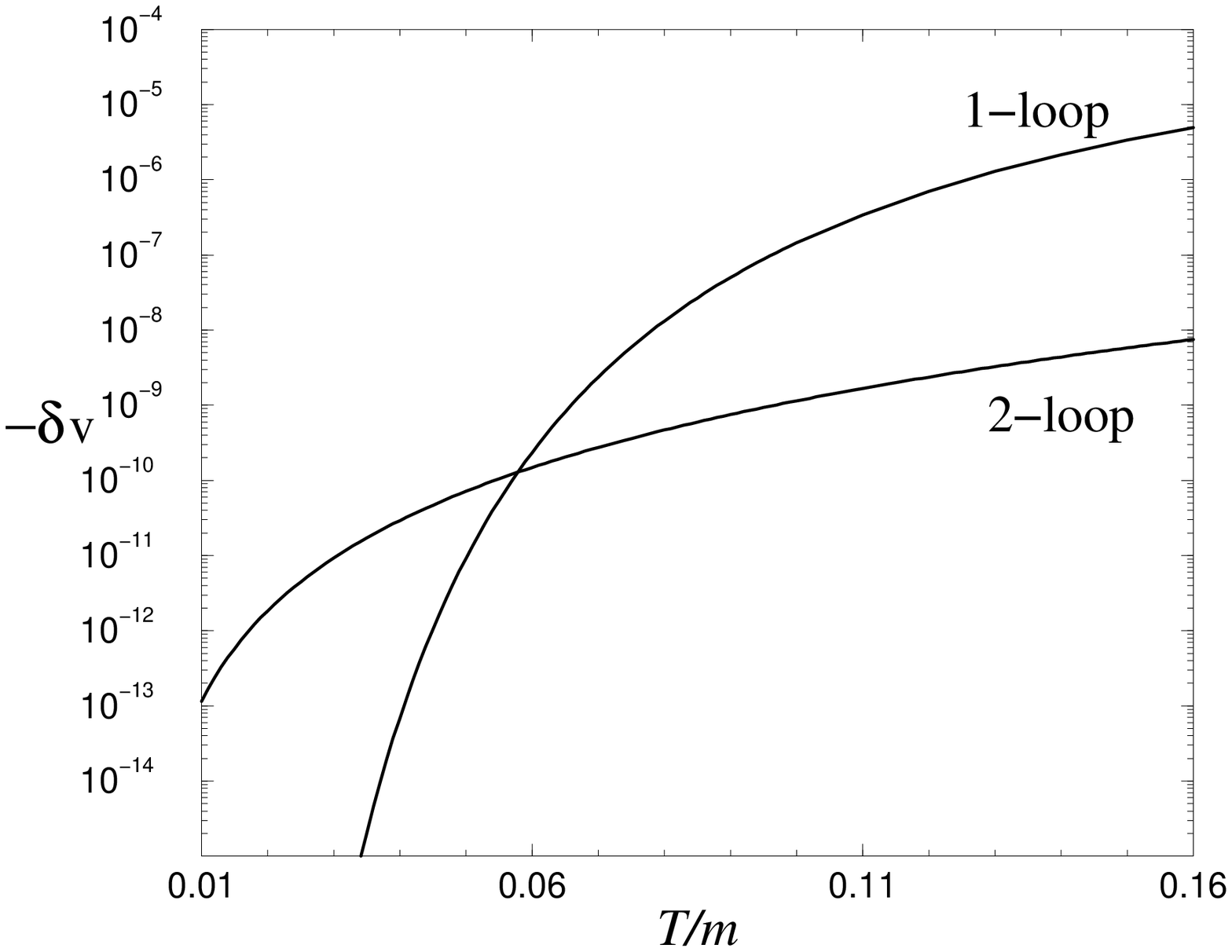,width=7.1cm}
}
\put(0,56){(a)} 
\put(85,0){
\epsfig{figure=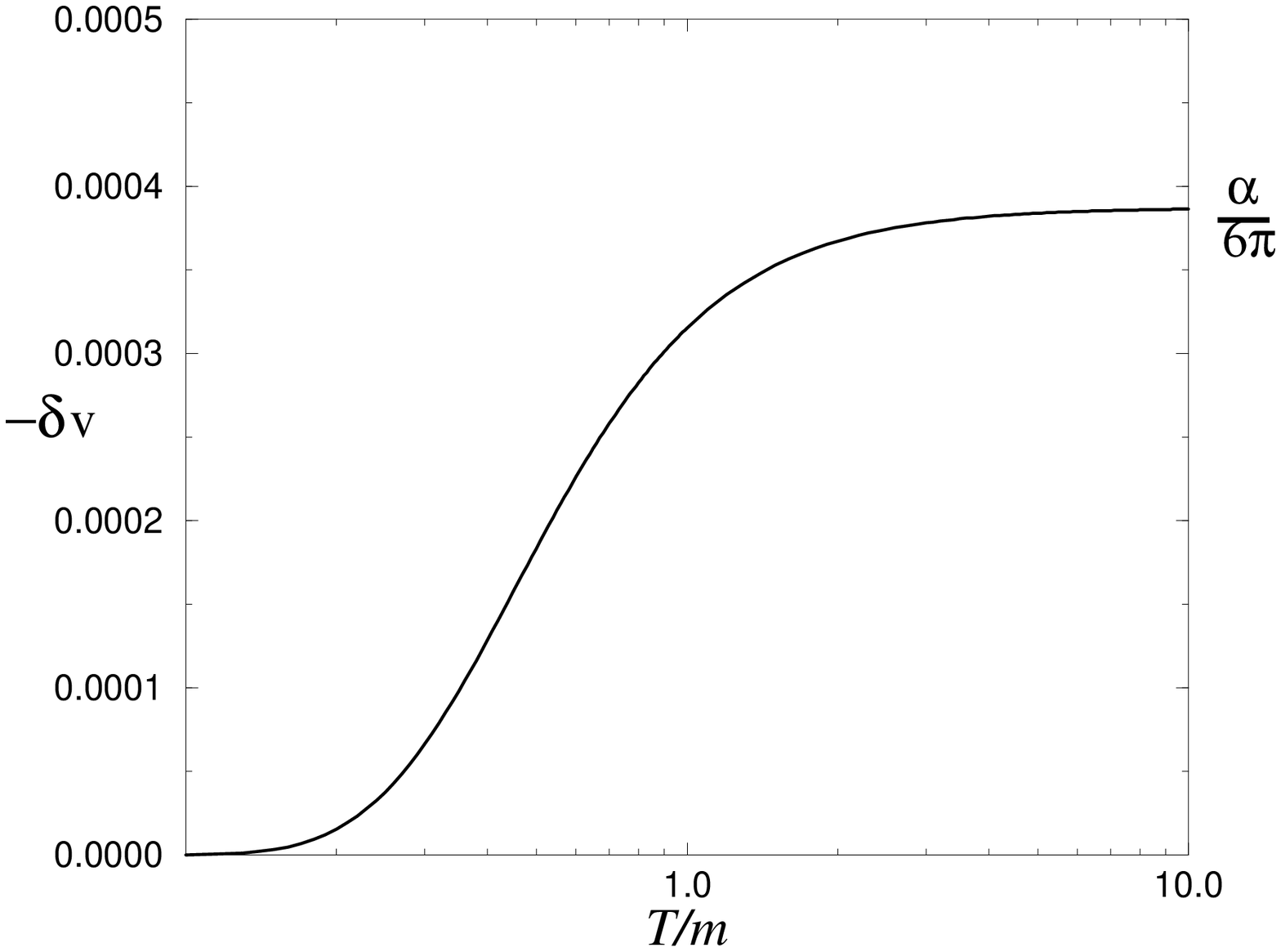,width=7.8cm}
}
\put(82,56){(b)}
\end{picture}
\end{flushleft}
 
\vspace{-.8cm}

\caption{(a) Low-temperature velocity shift $\delta v$ in units of the
  vacuum velocity $c=1$ versus temperature $T/m$; the one-loop 
  contribution exceeds the two-loop result at $T/m\simeq 5\%$. (b)
  Thermally induced velocity shift $\protect{-{\delta} v}$ in the
  intermediate-temperature domain $T\sim m$. The maximum velocity
  shift $-\delta v=\frac{\alpha}{6\pi}$ is already approached at
  comparably low temperatures.}
\label{figLP}
\end{figure}

Obviously, these velocity shifts are extremely difficult to measure;
perhaps one can make use of the fact that certain hot regions in the
universe such as stellar atmospheres or nebular structures can act as
lenses. Beyond that, it is interesting to note that, e.g., even in the
standard model of cosmology, the velocity of light has not been
constant during the evolution of the universe. Incidentally, velocity
shifts induced by finite $T$ {\em and} a magnetic field have also been
calculated; unfortunately, the combination of both vacuum
modifications leads rather to a washout of the single effects than to
an amplification.

Next, we turn to thermally induced photon splitting
\cite{Elmfors:1998ee,Gies:2000vb}, motivated by the fact that strong
magnetic fields and finite temperature (and density) may be
encountered near compact astrophysical objects. The new feature of the
thermal process is that the box graph with one coupling to the
background field (see Fig.~2(a)) is no longer suppressed by Lorentz
symmetry, since the latter is broken by the presence of the heat bath;
it is indeed the new invariant ${\cal E}$ which is responsible for the
lowest-order contribution to the thermal absorption coefficient of the
splitting process ($\bot \to \|_1+\|_2$):
 \begin{equation}
\frac{\kappa^T}{m} =\frac{1}{2^6\cdot3\cdot 5\pi^2}
\left(\frac{eB}{m^2}\right)^2 \sin^2\theta_B 
\left( \frac{\omega}{m}\right)^5 (\partial_{{\cal E}{\cal
      F}}{\cal L})^2 \,m^8, \label{PS20}
\end{equation}
where $\partial_{{\cal E}{\cal F}}{\cal L}$ denotes the coefficients 
of those terms in Eqs.\re{2Lft75} and \re{1124add1} which are bilinear
in ${\cal E}$ and ${\cal F}$. Note the different dependence on $B$
arising from the box graph contrary to Eq.\re{4}.

Soft photons with frequencies below the pair-production threshold are
also exposed to further absorption processes: first, the photons can
directly scatter with the $e^+e^-$ plasma; the absorption coefficient
for this Compton process can be estimated to give \cite{Elmfors:1998ee}
\begin{equation}
\frac{\kappa_{\text{C}}}{m} \simeq
\frac{8\alpha^2}{3\pi m^3}
 \int\limits_0^\infty dp \frac{p^2}{\E^{\omega_e/T}+1},
\label{PS25}
\end{equation}
where $\omega_e$ denotes the fermion energy
$\omega_e=\sqrt{p^2+m^2}$. The second competing process is the
scattering between the propagating photon and the blackbody radiation
of the thermal heat bath, i.e., light-by-light scattering as discussed
above. This absorption coefficient reads \cite{Gies:2000vb}:
\begin{equation}
\frac{\kappa_{\gamma\gamma}}{m} =\frac{7\cdot 139}{2^5\cdot 3^7\cdot
  5^6} \frac{\pi^9}{\zeta(3)^2} \alpha^4 \left(\frac{T}{m}\right)^6
  \left(\frac{\omega}{m}\right)^3 \simeq 5.21\cdot 10^{-11}
  \left(\frac{T}{m}\right)^6   \left(\frac{\omega}{m}\right)^3, \quad
  T\ll m. \label{gg4}
\end{equation}
These absorption coefficients have been depicted in Fig.~4 for typical
values of the parameters. The two-loop contribution dominates over the
one-loop contribution for $T/m\leq 0.041$. However, Compton scattering
is the dominant absorption in the intermediate-temperature domain,
$T\sim m$, and above; whereas $\gamma\gamma$ scattering with the
heat-bath photons can become the most important process in the
low-temperature domain for weak fields. The latter processes do not
contribute to the generation of polarized light, but rather wash out
the anisotropies stemming from photon splitting; therefore, the actual
value of the temperature or magnetic field of a pulsar might be
inferred from a measurement of the anisotropies. Since the
photon-splitting amplitudes have to be added coherently, a
hypersurface exists in parameter space where the amplitudes interfere
totally destructively. In actual measurements, this effect might be
visible as a sharp inverse peak in a plot of the polarization
asymmetries.

\begin{figure}
\begin{flushleft}
\begin{picture}(145,67)
\put(0,0){
\epsfig{figure=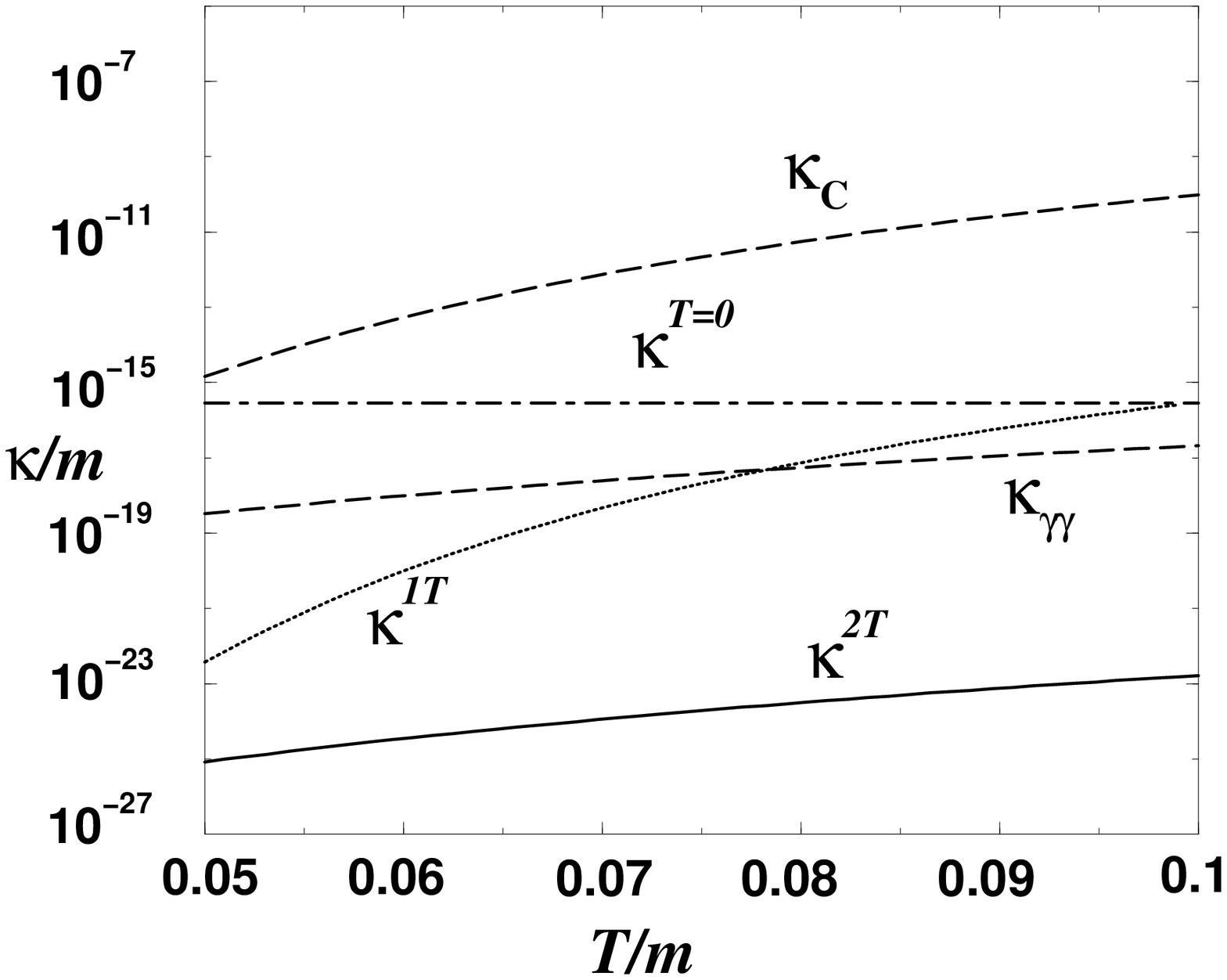,width=7.8cm}
}
\put(0,67){(a): \qquad$\frac{eB}{m^2}=0.2$, $\frac{\omega}{m}=1=\sin
  \theta_B$} 
\put(83,0){
\epsfig{figure=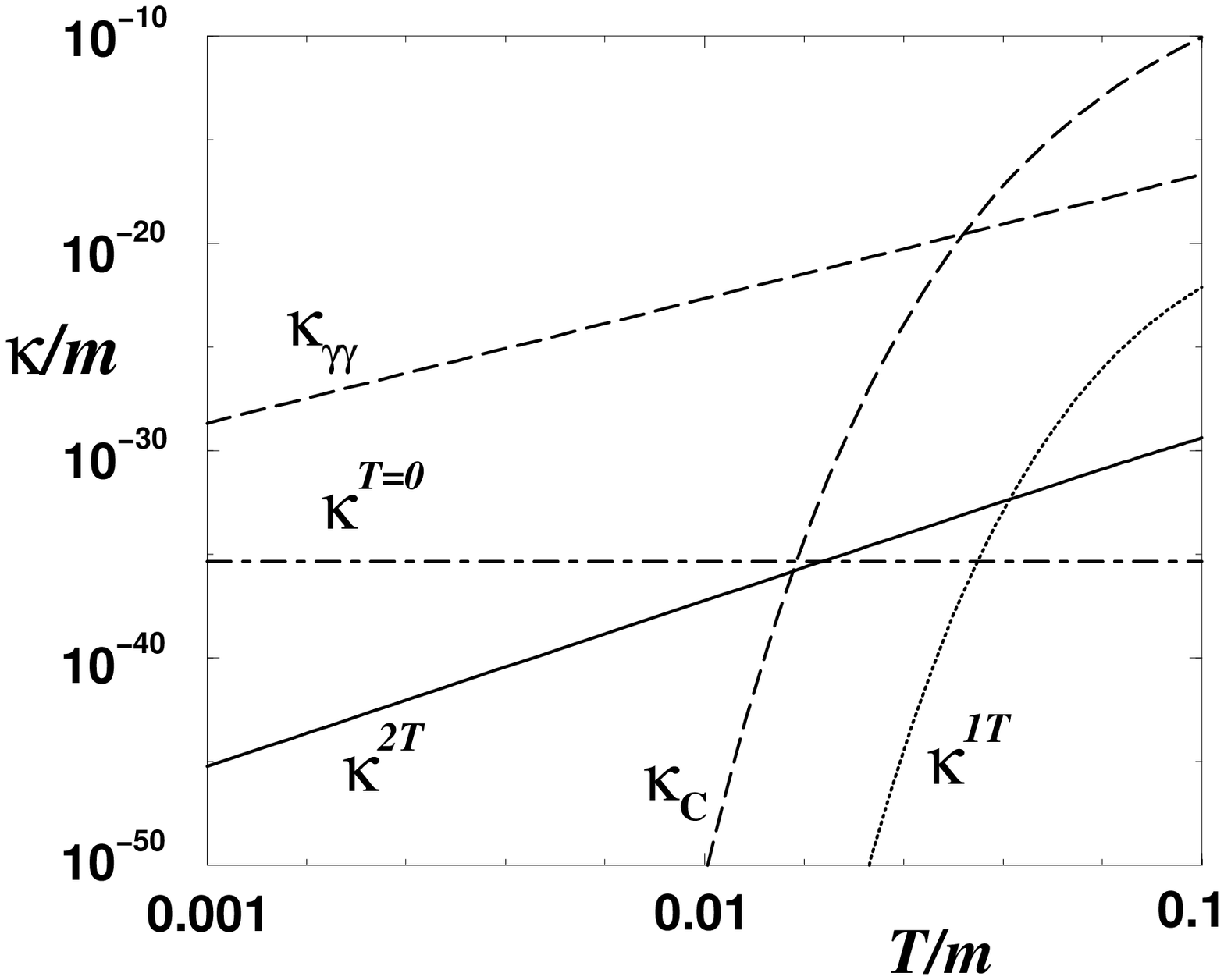,width=7.8cm}
}
\put(85,67){(b):\qquad$\frac{eB}{m^2}=10^{-4}$, $\frac{\omega}{m}=1=\sin
  \theta_B$}
\end{picture}
\end{flushleft}

\vspace{-.8cm}

\caption{Absorption coefficient $\kappa$ in units of the electron mass
  versus temperature $T$ in units of the electron mass. In Fig. (a),
  the various contributions are plotted for parameter values of a
  realistic astrophysical system. In Fig. (b), the parameters are
  chosen in such a way that the two-loop dominance over the one-loop
  and the Compton process is revealed; the photon-photon scattering
  contribution cannot be surpassed in the low-temperature limit.}
\label{figPS1}
\end{figure}

Finally, we turn to thermal contributions to the Schwinger mechanism
of pair production in electric fields. Combining the classical
zero-temperature result \cite{Schwinger:1951nm} with the leading
thermal two-loop contribution \cite{Gies:2000vb}, we find in the
low-temperature domain, $T\ll m$:
\begin{eqnarray}
\text{Im}\,\Leff(eE\!\ll\! m^2)\!\!\! &=&\!\! m^4 \E^{-\pi/\eta} \left
  (\! \frac{\eta^2}{8\pi^3} +\frac{\alpha\pi^2}{180} \frac{1}{\eta^2}
  \frac{T^4}{m^4}\! \right)\! \simeq m^4 \E^{-\pi/\eta} \left(\!\! 4\cdot
  10^{-3} \eta^2 +4\cdot10^{-4} \frac{T^4/m^4}{\eta^2}\!
  \right)\nonumber\\
\text{Im}\,\Leff(eE\!\gg\! m^2)\!\!\! &=&\!\!m^4 \,\eta\left(
  \frac{\eta}{48\pi} + 
  \frac{\alpha\pi}{270} \frac{T^4}{m^4} \right) \simeq m^4\, \eta
  \left( 6.6\cdot 10^{-3}\eta+8.5\cdot 10^{-5} \frac{T^4}{m^4}
  \right), \label{90z}
\end{eqnarray}
where we abbreviated $\eta=\frac{eE}{m^2}$. These equations show that
the low-temperature contribution to $\text{Im}\,\Leff$ can be
neglected for strong electric fields; the physical reason for this
lies in the fact that the nonlinearities of pure (zero-$T$) vacuum
polarization exceed the polarizability of the thermally induced real
plasma by far in the strong field limit. By contrast, in the limit of
weak electric fields, thermal effects can increase the pair-production
probability significantly. Of course, for these values of $\eta$, the
total imaginary part is very small due to the inverse power of $\eta$
in the exponential. 

\section{Effective action for photon-neutrino interactions}

As mentioned above, if the electron had a much lighter sister, we
would experience our world quite differently, because (classical)
electrodynamics would be inherently nonlinear. In fact, there are
very light fermions, the neutrinos, but, (un-)fortunately, they do not
couple to electromagnetic fields in a sufficiently strong way. But they
nevertheless couple to these fields via intermediary electrons and
gauge bosons. Therefore, if we are interested in the physics below the
electron mass scale, neutrinos have to be taken into consideration in
addition to soft electromagnetic fields. In particular, we are looking
for an effective action governing the dynamics of soft photons
interacting with neutrinos. 

Since we only want to consider energies far below the heavy gauge
boson masses, we may start from another effective theory, the Fermi
theory, describing the four-fermion interaction between neutrinos and
the leptons\footnote{In the following, we only consider the influence
  of electrons; the heavier mass of the other leptons (and quarks)
  strongly suppresses their additional effects.}. For soft photons and
``soft'' neutrinos $\nu$, we arrive at their effective theory by
integrating out (averaging over) the actual ``heaviest'' particle,
i.e., the electron, and find
\begin{equation}
\Geff=\frac{\GF}{\sqrt{2}} \frac{1}{e}\int d^4x\,
\overline{\nu }\gamma ^{\mu } \left( 1+\gamma _{5}\right) \nu\,\, 
\bigl( g_{\text{V}} \langle j^\mu\rangle^A +g_{\text{A}} \langle
j_5^\mu\rangle^A \bigr) ,\label{Leff1}
\end{equation}
where we introduced the expectation values of the electronic vector
and axialvector current in a background field
$A_\mu$. Diagrammatically, Eq.\re{Leff1} describes an electron loop in
a background field with a $\nu\bar{\nu}$ insertion as depicted in
Fig.~5. $\GF$ denotes the Fermi constant. 

The vector current expectation value can easily be obtained using the
well-known Heisenberg-Euler action \re{1} and the formula $\langle
j_\mu\rangle^A=-\delta \Geff^{HE}/\delta A^\mu$. In this way, one
obtains a derivative expansion of the vector current around a strong
field.  For example, the term which is third order in the field and
first order in derivatives was used by Dicus and Repko
\cite{Dicus:1997rw} in their study of $\nu \gamma \to \nu \gamma
\gamma$ and cross processes (see also \cite{Abada:1999sw}), and by
Shaisultanov \cite{Shaisultanov:1998bc} for investigating the
$\nu\gamma\to\nu\gamma$ and crossed processes in a background magnetic
field. Unfortunately, such a simple formula for the axialvector
current $\langle j_5^\mu\rangle$ does not exist, so it must be
calculated from first principles. This has been achieved recently in
\cite{Gies:2000tc} for arbitrarily strong electromagnetic background
fields in a first-order gradient expansion. The necessary basic
equation is the relation between the axialvector current and the
axialvector-vector amplitude, i.e., the axial analogue of the
polarization tensor,
\begin{equation}
\langle j_5^\mu\rangle^A=\frac{1}{3}\partial_{\sigma}F_{\alpha
  \beta}\frac{\partial^2}{\partial k^{\sigma}\partial k^{\alpha}}
\Bigl[\Pi_5^{\beta \mu}(-k)\Bigr]\Big|_{k=0},
\label{ax}
\end{equation}
which holds to first order in derivatives.\footnote{A similar equation
  holds also for the vector current and the polarization tensor.} The
axialvector-vector amplitude has been calculated independently and
with very different methods in \cite{Shaisultanov:2000mg} and
\cite{Schubert:2000kf}. Here, we shall be satisfied by citing the final
  result for the axialvector current in weak-field expansion:
\begin{eqnarray}
\langle j_{5}^{\mu }\rangle ^{A}&=&\frac{\displaystyle
e^{3}}{\displaystyle 24\pi ^{2}m^{2}}
\Bigl (\partial^{\mu} {\cal G}+(\partial ^{\alpha }F_{\alpha \beta })
F^{*\beta \mu }\Bigr ) \label{weakf}\\
&&+\frac{\displaystyle e^{5}}{\displaystyle 90\pi ^{2}m^{6}}
\partial _{\sigma }F_{\alpha \beta }
\Bigl [{\cal G} \bigl(F^{\beta \alpha }g^{\mu \sigma }+
F^{\beta\sigma} g^{\mu\alpha}\bigr)
+ \bigr( F^{*\beta \alpha }(F^{2})^{\mu \sigma }
     + F^{*\beta\sigma} (F^2)^{\mu\alpha}\bigr) \nonumber\\
&&\qquad\qquad\qquad\qquad - {\cal F}\bigl(3F^{*\mu \beta }g^{\alpha
  \sigma }+  F^{*\beta \alpha }g^{\mu \sigma }+
F^{*\beta \sigma }g^{\mu \alpha }\bigr)\Bigr ]. \nonumber 
\end{eqnarray}
Incidentally, the first term can also be obtained from the famous
triangle diagram \cite{Gies:2000wc}. Upon insertion of Eq.\re{weakf}
and the well-known vector current into Eq.\re{Leff1}, we arrive at the
analogue of the Heisenberg-Euler action for the case involving an
axialvector coupling. This provides us with the effective action for
neutrino interactions with an arbitrary soft electromagnetic field.

\begin{figure}
\begin{center}
\begin{picture}(70,40)
\put(5,0){
\epsfig{figure=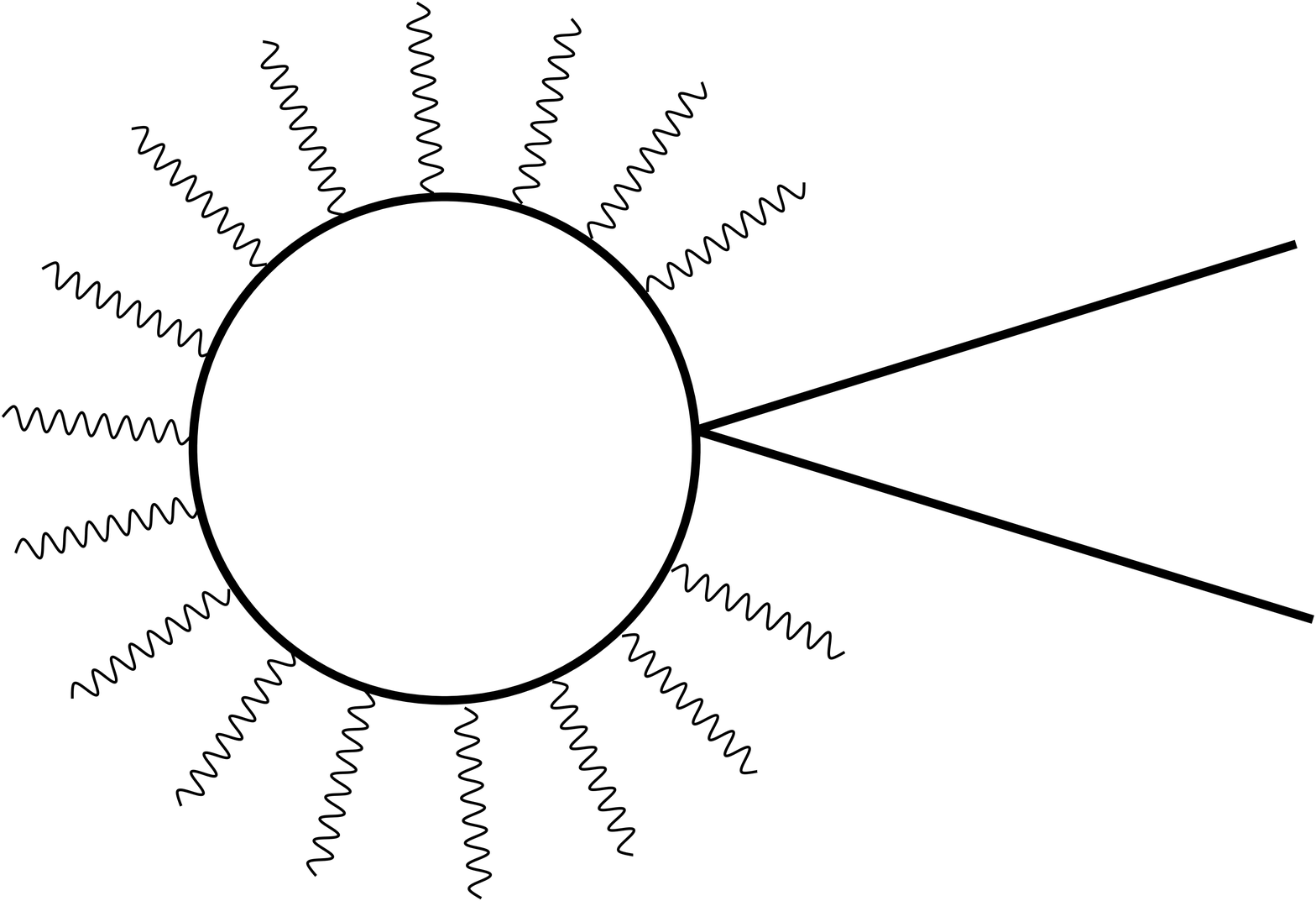,width=6.5cm}}
\put(65,33){$\bar{\nu}$}
\put(65,11){$\nu$}
\end{picture}
\end{center}

\vspace{-0.8cm}

\caption{Diagrammatic representation of the effective action for soft
  electromagnetic fields coupling to neutrinos.} 
\end{figure}

The application of this effective action to real physical systems has
not been fully explored up to now, and we shall sketch only a few
examples, concentrating on the axialvector-coupling sector. The first
example resembles to some degree the photon-splitting process:
Cherenkov radiation of neutrinos in magnetic fields. Since the
velocity of light is reduced in magnetic fields (cf. Eq.\re{6}),
neutrinos can propagate ``faster than light'' and emit Cherenkov
radiation. For massless neutrinos propagating perpendicular to the
magnetic field, the rate for producing collinear Cherenkov radiation
with $\|$ polarization WRT the magnetic field is \cite{DeRaad:1976kd}
\begin{equation}
\Gamma_{\nu\to\nu\gamma}(m_\nu=0)=\frac{16\alpha \GF^2}{3^3\cdot
  5^3\cdot(4\pi)^4}\, E^5 \left(\frac{B}{\Bcr}\right)^6, \quad B\ll
  \Bcr,\label{11} 
\end{equation}
where $E$ is the incident neutrino energy. The result is independent
of neutrino flavors. Similarly to the case of photon splitting (at
$T=0$), we observe a $B^6$ dependence signifying three couplings of
the background field to the electron loop as the lowest-order
contribution; together with the neutrino vertex and the outgoing
photon, this corresponds to a pentagon diagram, represented by the
last two lines in Eq.\re{weakf}. The triangle diagram is suppressed
for symmetry reasons, similarly to the box diagram in photon
splitting.

This suppression no longer holds if we take nonvanishing neutrino
masses $m_\nu$ into account. Then the dominant mass-dependent
contribution arises from the first line of Eq.\re{weakf}, yielding
\cite{Gies:2000wc}
\begin{eqnarray}
\Gamma_{\nu\to\nu\gamma}(m_\nu\!\neq0)\!\!\!&=&\!\!\!
\frac{7}{2^9\cdot3^4\cdot 5^2}
\frac{\alpha^2}{\pi^4} \, m \, (\GF m^2)^2 
\left(\!\frac{E}{m}\!\right)^3\!\left(\! 1-\frac{E_{\text{min}}^2}{E^2}
\right)^5\! \left( \!\frac{m_\nu}{m}\!\right)^2\! \left( \!
  \frac{B}{\Bcr}\! \right)^4 \theta(E-\!E_{\text{min}}), \nonumber\\
&&\text{where}\quad E_{\text{min}}:= \sqrt{\frac{45\pi}{7\alpha}} m_\nu
\frac{\Bcr}{B}.\label{12}
\end{eqnarray}
The existence of such a mass-dependent threshold energy
$E_{\text{min}}$ for the incoming neutrino arises from the Cherenkov
condition: the neutrino must move ``faster than light'' in the
$B$-field background. Astonishingly, the transition rate \re{12} never
wins out over the mass-independent contribution of Eq.\re{11}. This
nontrivial zero-result should serve as a counterexample for the
general belief that non-zero neutrino masses always open a window in
parameter space for new effects.

As as second example, let us consider $\bar{\nu}\nu$ pair emission in
varying electromagnetic fields \cite{Gies:2000wc}. The pure
QED-analogue of $e^+e^-$ pair emission in such fields, though it is
textbook knowledge, remains phenomenologically unimportant because of
the ``heavy'' mass of the electron. The similar mechanism for
neutrinos benefits from the smallness of the neutrino mass, and
originates from the first term of Eq.\re{weakf} as a part of the
triangle diagram.

This term gives rise to the matrix element ${\cal M}({\cal G}(k) \to
\bar{\nu}(p'),\nu(p))$, where ${\cal G}(k)$ denotes the Fourier
transform of the pseudoscalar invariant ${\cal G}(x)$ of the
electromagnetic field. Squaring the matrix element and integrating
over the phase space, we obtain the production probability
\begin{equation}
W=\frac{\GF^2 \alpha^2}{36 (2\pi)^7} \frac{m^2_\nu}{m^4} \int d^4k \,
|{\cal G}(k)|^2\, k^2\left( 1- \frac{4m_\nu^2}{k^2} \right)^{1/2}
\theta\left(1- \frac{4m_\nu^2}{k^2} \right). \label{13}
\end{equation}
In order to obtain an illustrative estimate of the order of magnitude
of this effect, let us simply take ${\cal G}(k)= \mathbf{E_0\cdot B_0}
(2\pi)^4 \delta^3(\mathbf{k}) \delta(\omega -2\omega_0)$. For this
field configuration, we obtain the production probability per volume
and time
\begin{equation}
\frac{W}{V T} \simeq 5.11
 \left( \frac{m_\nu}{1\text{eV}}\right)^2
 \left(\frac{\omega_0}{1\text{eV}} \right)^2 \frac{(\mathbf{E_0\cdot
 B_0})^2}{\Bcr^4} \left(1-\frac{m_\nu^2}{\omega_0^2}\right)^{1/2}
 \theta (\omega_0-m_\nu)\, {\text{cm}^3}{\text{s}}, \label{14}
\end{equation}
in units of cm${}^3$ and seconds. Obviously, the threshold frequency
is equal to the neutrino mass, e.g., in the strong optical ultraviolet
for neutrino masses at the eV scale.  Equation \re{14} can also be
interpreted as the number of pairs produced in the system and volume
under consideration.

It is instructive to compare this pair-production
probability with the one for $e^+e^-$ pairs: $W_{e^+e^-}\sim \int
(E^2-B^2)$. Each process is triggered by a different invariant of the
electromagnetic field revealing its vector or axialvector character. 

Except for mass differences, the above-discussed examples are not
directly sensitive to different neutrino flavors. This is different
in the following example: consider a spatially constant
electromagnetic vacuum field varying in time with non-vanishing
$\mathbf{E\cdot B}\neq 0$. Its contribution to the axial charge
density is $\frac{\gA}{e} \langle j_5^0\rangle
=\frac{\alpha}{6\pi}\frac{\gA}{m^2} \frac{d}{dt} \mathbf{E\cdot B}$.
Therefore, such a field configuration is in complete analogy to a
polarized medium. In this way, a neutrino propagating in such a field
can be subject to an enhancement of flavor oscillations, similarly to
a propagation in matter. \bigskip

This concludes our selection of QED effective actions which govern the
physics of soft electromagnetic fields and thereby connect features of
the quantum vacuum with classical electrodynamics. The variety of
physical effects, partly discussed here and partly still to be
discovered, are manifold but share the common feature that their
experimental verification will require extraordinary abilities and
facilities. Nevertheless, their investigation is worthwhile and will
deepen our understanding of the quantum vacuum.

\section*{Acknowledgement}

I would like to thank W.~Dittrich and R.~Shaisultanov for many
insightful discussions and collaboration on some of the topics
presented here. W.~Dittrich's thoughtful reading of the manuscript is
also gratefully acknowledged. Finally, it is a pleasure to thank
G.~Cantatore and the other organizers of this productive and enjoyable
workshop for their efforts.


\begin{thebibliography}{99}
{\small
\setlength{\itemsep}{-.2mm}
\bibitem{Dittrich:2000zu}
W.~Dittrich and H.~Gies,
``{\em Probing the quantum vacuum. Perturbative effective action
  approach in quantum electrodynamics and its application}'', 
Springer Tracts Mod.\ Phys.\  {\bf 166}, 1 (2000).
\bibitem{heis} H. Euler and B. Kockel, Naturwissenschaften
  {\bf 23}, 246 (1935);  W. Heisenberg and H. Euler, Z. Phys. {\bf
  98}, 714 (1936); V. Weisskopf,
  K. Dan. Vidensk. Selsk. Mat. Fy. Medd. {\bf 14}, 1 (1936). 
%
\bibitem{Schwinger:1951nm}
J.~Schwinger,
Phys.\ Rev.\  {\bf 82}, 664 (1951).
%
\bibitem{Adler:1971wn}
S.~L.~Adler,
Annals Phys.\  {\bf 67}, 599 (1971).
%
\bibitem{peng98} R.~Pengo et al.; W.-T.~Ni, in {\em Frontier Tests of
    QED and Physics of the Vacuum}, Edited by E.~Zavattini, D.~Bakalov
  and C.~Rizzo, Heron Press Sofia (1998); E.~Zavattini; C.~Rizzo, this
    volume (2000).
%
\bibitem{Dittrich:1998fy}
W.~Dittrich and H.~Gies,
Phys.\ Rev.\  {\bf D58}, 025004 (1998)
[hep-ph/9804375];
Phys.\ Lett.\  {\bf B431}, 420 (1998)
[hep-ph/9804303].
W.~Dittrich and H.~Gies,
in {\em The Casimir Effect 50 Years Later}, Proceedings of
the Fourth Workshop on Quantum Field Theory under the Influence of
External Conditions, Edited by M. Bordag, World Scientific, (1998)
[hep-ph/9903469].
%
\bibitem{Novello:2000pg}
M.~Novello, V.~A.~De Lorenci, J.~M.~Salim and R.~Klippert,
Phys.\ Rev.\  {\bf D61}, 045001 (2000)
[gr-qc/9911085]; 
S.~Liberati, S.~Sonego and M.~Visser,
quant-ph/0010055.
\bibitem{bial70} Z. Bia{\l}ynicka-Birula and
  I. Bia{\l}ynicki-Birula, Phys. Rev. D {\bf 2}, 2341 (1970).
%
\bibitem{Dittrich:1998gt}
W.~Dittrich and H.~Gies,
 in {\em Frontier Tests of QED and Physics of the Vacuum}, Edited by
  E. Zavattini, D. Bakalov and C. Rizzo, Heron Press Sofia (1998)
[hep-ph/9806417].
%
\bibitem{Burke:1997ew}
D.~L.~Burke {\it et al.},
Phys.\ Rev.\ Lett.\  {\bf 79}, 1626 (1997);
A.~C.~Melissinos,
hep-ph/9805507.
%
\bibitem{Elmfors:1995fw}
P.~Elmfors and B.~Skagerstam,
Phys.\ Lett.\  {\bf B348}, 141 (1995)
[hep-th/9404106].
%
\bibitem{Gies:1999vt}
H.~Gies,
Phys.\ Rev.\  {\bf D60}, 105002 (1999)
[hep-ph/9812436].
%
\bibitem{Elmfors:1998ee}
P.~Elmfors and B.~Skagerstam,
Phys.\ Lett.\  {\bf B427}, 197 (1998)
[hep-ph/9802397].
%
\bibitem{Gies:2000vb}
H.~Gies,
Phys.\ Rev.\  {\bf D61}, 085021 (2000)
[hep-ph/9909500].
%
\bibitem{Gies:1999xn}
H.~Gies,
Phys.\ Rev.\  {\bf D60}, 105033 (1999)
[hep-ph/9906303].
%
\bibitem{Barton:1990dq}
G.~Barton,
Phys.\ Lett.\  {\bf B237}, 559 (1990);
R.~Tarrach,
Phys.\ Lett.\  {\bf B133}, 259 (1983);
M.~H.~Thoma,
hep-ph/0005282.
%
\bibitem{Dicus:1997rw}
D.~A.~Dicus and W.~W.~Repko,
Phys.\ Rev.\ Lett.\  {\bf 79}, 569 (1997)
[hep-ph/9703210].
%
\bibitem{Abada:1999sw}
A.~Abada, J.~Matias and R.~Pittau,
Phys.\ Rev.\  {\bf D59}, 013008 (1999)
[hep-ph/9806383];
Y.~Aghababaie and C.~P.~Burgess,
hep-ph/0006165.
%
\bibitem{Shaisultanov:1998bc}
R.~Shaisultanov,
Phys.\ Rev.\ Lett.\  {\bf 80}, 1586 (1998)
[hep-ph/9709420].
%
\bibitem{Gies:2000tc}
H.~Gies and R.~Shaisultanov,
Phys.\ Rev.\  {\bf D62}, 073003 (2000)
[hep-ph/0003144].
%
\bibitem{Shaisultanov:2000mg}
R.~Shaisultanov,
hep-th/0002079.
%
\bibitem{Schubert:2000kf}
C.~Schubert,
Nucl.\ Phys.\  {\bf B585}, 429 (2000)
[hep-ph/0002276].
%
\bibitem{Gies:2000wc}
H.~Gies and R.~Shaisultanov,
Phys.\ Lett.\  {\bf B480}, 129 (2000)
[hep-ph/0009342].
%
\bibitem{DeRaad:1976kd}
L.~L.~DeRaad, K.~A.~Milton and N.~D.~Hari Dass,
Phys.\ Rev.\  {\bf D14}, 3326 (1976);
A.~N.~Ioannisian and G.~G.~Raffelt,
Phys.\ Rev.\  {\bf D55}, 7038 (1997)
[hep-ph/9612285].
}
\end{thebibliography}
\end{document}